\documentstyle[12pt]{article}
\title{
 Hybrid  states  with  heavy  quarks
}
\vspace{2cm}
\author{
Yu. B. Yufryakov\thanks
{ e-mail : yufryakov@vxitep.itep.ru}
\\ Institute of Theoretical and Experimental Physics\\ 117259,
Moscow}
\date{}
\newcommand{\be}{\begin{equation}}
\newcommand{\ee}{\end{equation}}
\begin{document}
\maketitle

\begin{abstract}

  Model-independent  features  of
    hybrid states  with  heavy
      quarks and  adiabatic  approximation
       for  such  systems  are  discussed.
        These  general arguments are  applied
          to  the  QCD-based  constituent  gluon  model.
 The  leading-order  Hamiltonian  is  derived  from
   the  general  formalism.  The  masses  of
     heavy quarks  hybrids, including   charmonium  hybrid  states,
     are  calculated. We  predict $c\bar c g$
     hybrids at 4.0$\pm$0.1
      GeV  and
     $b \bar b g$
      ones at 10.5$\pm$0.1 GeV.
       Comparison with  some other models
      are  made.
  Model-independent  features  of
    hybrid states  with  heavy
      quarks and  adiabatic  approximation
       for  such  systems  are  discussed.
        These  general arguments are  applied
          to  the  QCD-based  constituent  gluon  model.
 The  leading-order  Hamiltonian  is  derived  from
   the  general  formalism.  The  masses  of
     heavy quark  hybrids, including   charmonium  hybrid  states,
     are  calculated. We  predict $c\bar c g$
     hybrids at 4.0$\pm$0.1
      GeV  and
     $b \bar b g$
      ones at 10.5$\pm$0.1 GeV.
       Comparison with  some other models
      are  made.
\end{abstract}

      \newpage

     The  problem  of   exotic  states with  heavy  quarks  is
     widely  discussed  in  the  literature. Despite  the  fact
     that  there  is  no  unambigous  evidence  of  existence  any
     exotics  even  for  light  quarks  sector, intensive  theoretical
     studies  of  heavy  quarks  exotics  are  carrying  out(see [5]
     for review). The
     reason  of  this  interest  is  that  the heavy  flavours
     spectroscopy  is  well   understood, much  more  better  than
     for  light  quarks  sector. Presence  of  the  slow
     nonrelativistic  quark  subsystem  allows  one  to  use  the
     adiabatic  (Born-Oppenheimer)  approximation.

         So, there are
        slow  quark  subsystem and  fast  gluonic  subsystem .  The
          only  mass  parameter for  gluonic subsystem is  string
          tension  $\sigma$. Its   energy
          in the order  of  magnitude   is
          $E_{gl}\propto \sqrt{\sigma}$;
          radius is
          $r_{gl}\propto 1/\sqrt{\sigma}$;
              and  velocity  is
          $\dot r_{gl}\propto 1$.
           Assume  that  quarks  are  confined
          by oscillator  potential  produced  by  gluonic  subsystem
          .  Then  for  quark  subsystem  we  can  estimate :
\be
E_{q \bar q}\propto \sigma^{3/4}/\sqrt{m};
          ~~~r_{q \bar q }\propto 1/~^4\sqrt{m\sigma^{3/2}}
          \ee

          Hence , ratio  of
              the periods  of  slow  and  fast  subsystems  equals :
      \be
      T_{q\bar q}/T_{gl}\propto E_{gl}/E_{q\bar q}\propto
      \sqrt{m/\sqrt{\sigma}}
      \ee
                           and radii ratio is :
      \be
      r_{q\bar q}/r_{gl}\propto
      ~^4\sqrt{\sqrt{\sigma}/m}
      \ee

 It's   easy  to  see that  a    typical small
       parameter  is    $\sqrt{\sqrt{\sigma}/m}$.
       Unfortunately, because of the numerical reasons estimation ( 3 )
       is  not satisfied for c-  and b-quarks . But the main point we
       stress is  that the oscillator potential gives  us   quark
       subsystem energy  of order
       $\sqrt{\sigma}\sqrt{\sqrt{\sigma}/m}$
        and  any
       unharmonic corrections  will be  of  the next  orders.

       So ,
       we propose  the  following description  of hybrids  with
        heavy quarks . Suppose  we have  a  Hamiltonian for
        $q \bar q g$--
         system .  We'll  obtain the leading
        approximation in the  limit
        $r_{q\bar q}\to 0$.
         We can
        calculate  gluon  subsystem energy solving Schrodinger-type
        equation with  our Hamiltonian . To  obtain quark subsystem
        energy  we  ought to 1) substract  from  our Hamiltonian  all
        terms proportional  to  the  square  of the interquark
        radius.  2)  Averaged over gluon  wave function these  pieces
        of the  full  Hamiltonian produce us Hamiltonian  of  the
        quark  subsystem   which in  turn gives us  quark corrections
        to  the   energy  of  the gluonic subsystem .

            Our  starting
        point  is  the constituent  gluon model described  in [1]  .
        Using Feynman-Schwinger representation for Green function one
       can write effective Lagrangian for $q\bar q g$
       system in the Euclidean space-time
       (spin degrees of freedom are neglected):
       \be
       L=\frac{2\mu+\mu_g}{2} +\frac{m^2}{\mu}+\frac{\mu\dot{z}^2}{2}+
       \frac{\mu\dot{\bar{z}}^2}{2}+\frac{\mu_g\dot x^2}{2}+
       \sigma{r_1}\int^1_0d\beta_1\sqrt{1+l^2_1}+\sigma{r_2}\int^1_0
       d\beta_2\sqrt{1+l^2_2}
       \ee
       $$
       l_1^2=(\beta_1\dot{z}+(1-\beta_1)\dot
       x)^2-\frac{1}{r^2_1}(\beta_1(\dot z{r_1})+(1-\beta_1)(\dot
       z{r_1}))^2 $$
        $$ l_2^2=(\beta_2\dot{\bar z}+(1-\beta_2)\dot
       x)^2 -\frac{1}{r^2_2}(\beta_2(\dot{\bar z}{r_2})+(1-\beta_2)
       (\dot{\bar z}{r_2}))^2 $$

        (We consider
 equal quark masses case, Coulomb interaction omitted).  Here
 $\mu$ and $\mu_g$ are effective  masses of quark and
 gluon respectively ; $z,\bar{z}$ and $x$ are  quark,
 antiquark and gluon 3-coordinates ;
 $r_1=z-x ; r_2=\bar z-x$.

 To deal  with the  square root terms we use  the auxiliary
 field formalism [2]. Following the method of [2] we obtain the
 Lagrangian:
 $$
 L=\frac{m^2}{\mu}+\mu+\frac{1}{2}[\mu_g(1+\dot x^2)+\mu(\dot z^2+
\dot{\bar z}^2)+(\int^1_0d\beta_1 \nu_1(\beta_1)(1+l^2_1)+
 $$
\be
 +\sigma^2r^2_1
 \int^1_0\frac{d\beta_1}{\nu_1(\beta_1)}+(1\to 2))]
\ee

   To
 eliminate
 center-of -mass motion we can introduce variables (analogously to
 the [2]) :
 \be
 r_{1i}=z_i-x_i;~~
 r_{2i}=\bar z_i-x_i;~~
 R_{i}=(1-\xi_1-\xi_2)x_i+\xi_1z_i+\xi_2\bar z_i
 \ee
 where $\xi_{\gamma} (\gamma = 1,2 )$ are parameters to be defined
 from  the condition that all terms  of the type
 $(\dot{R}\dot r_1)$ and
 $(\dot R \dot r_2)$
  cancel ( but  not
  $(\dot{R} r_1)$ $(\dot{r} r_1)$
  type terms!).  It's easy  to see  that

\be
\xi_1=\frac{\mu_1+\int^1_0 d\beta_1\nu_1 (\beta_1)\beta_1}{\mu_1+\mu_2
+\mu_g+\int d\beta_1\nu_1+\int d\beta_2\nu_2};~~
\xi_2=(1\leftrightarrow 2)
\ee
  After that  the integration over
 center-of-mass coordinate $\vec R$ can  be
 carried out .  The resulting Lagrangian reads :

 \newpage

 $$
 L=\frac{m^2}{\mu}+\frac{M}{2}+\frac{\dot r_1^2}{2}
 [\mu(1-2\xi_1+2\xi_1^2)+\mu_g   \xi^2_1+c_1+
 a_2\xi^2_1]+(1\leftrightarrow 2) - M\xi_1\xi_2 (\dot r_1
 \dot r_2)+
 $$
 $$+ \frac{d_1\sigma^2r^2_1}{2}-
 \frac{c_1}{2} \frac{(\dot r_1 r_1)^2}{r^2_1}+
 (1\leftrightarrow 2 )
 -\frac{1}{2A}\{\frac{1}{r^2_1}[b^2_1(\dot r_1
 r_1)^2(a_1+M+a_2\frac{(r_1r_2)^2}{r^2_1r^2_1})-
 $$
 $$
 -2\xi_2Mb_1(a_1+\eta)(\dot r_1 r_1)(\dot r_2 r_1)+
 \xi^2_2Ma_1(a_1+\eta)(\dot r_2
 r_1)^2]+
 $$
 \be
 +(1\leftrightarrow 2) +[b_1(\dot r_1
 r_1)-\xi_2a_1(\dot r_2r_1)][b_2(\dot r_2 r_2)-\xi_1
 a_2(\dot r_1 r_2)] \frac{2M(r_1r_2)}{r^2_1r^2_2}\}
 \ee
 where
 $$
 a=\int^1_0 d\beta \nu (\beta);~~ b=\int^1_0 d\beta \nu
 (\beta)(\beta-\xi); ~~ c=\int^1_0d\beta\nu
 (\beta)(\beta-\xi)^2; ~~ d=
 \int^1_0\frac{d\beta}{\nu(\beta)}
 $$
 \be
 \eta=\mu_1+\mu_2+\mu_g;~~ M=\eta+a_1+a_2;~~A=M\eta +
 a_1a_2\frac{|\vec r_1\times \vec r_2|^2}{r^2_1r^2_2}
 \ee

   We  assume that $\nu\sim \sqrt{\sigma}$. Expanding (8) on
 the $\sqrt{\sigma}/m$  and using new
 variables
 \be
 r_1=\frac{z_1+\bar z_1}{2}-x_1=\frac{r_{1i}+r_{2i}}{2};~~
 \rho_1=z_i-\bar z_i=r_{1i}-r_{2i}
 \ee
 it's easy to see that in the leading order relative
 quark motion (with reduced mass $\mu/2$ ) is
 decoupled :
 \be
 \frac{\mu}{4}(\dot r_1-\dot r_2)^2=\frac{\mu}{4}\dot{\rho}^2
\ee

   In the leading order we have  for gluonic subsystem :
 \be
 L_{gl}=\frac{\dot
 r^2}{2}(\mu_g+\Pi_1+\Pi_2)-\frac{\Pi_1}{2}\frac{(\dot r
 r_1)^2}{r_1^2}-\frac{\Pi_2}{2}\frac{(\dot r r_2)}{r_2^2}+
 \frac{1}{2}(\mu_g+a_1+a_2+\sigma^2r^2_1d_1+\sigma^2r^2_2d_2)
 \ee
  where
  \be
 \Pi=c-b+a/4=
 \int^1_0d\beta\nu(\beta)(\beta-1)^2
 \ee

   The kinetic term  in the Lagrangian (12) is
 quadratic in velocities.  Hence one can easily obtain Hamiltonian
 and all terms proportional to $\rho^2$
  can be
 separated.The  result reads :
 $$
\hat T_{gl}=\hat T^{(0)}+\hat T^{(1)}
 $$
 $$
 \hat
 T^{(0)}=\frac{1}{2\mu_g}
 \frac{(pr)^2}{r^2}+\frac{\hat{\vec{L}}_{gl}^2}{2(\mu_g+\Pi_1+\Pi_2)r^2}
 $$
 \be
 \hat T^{(1)}(\rho)=\frac{\Pi \rho^2}{6}[\frac{\hat{\vec
 L}^2_{gl}}{2(\mu_g+2\Pi)^2r^4}-\frac{(\vec p\vec n)^2}{\mu^2_g
 r^2}]
 \ee

  We
 assumed S-wave for quarks in ( 14 ) :
 \be
 <\rho_i\rho_j>=\frac{\rho^2\delta_{ ij}}{3}
 \ee

 Averaging (14) over chosen gluon wave function we obtain a
 contribution  to the oscillator potential for quark subsystem. In
 fact, the main contribution arise from the "potential" term of
 Lagrangian (12)  and (14)  is only a small correction.
 \par Zeroth
 approximation Hamiltonian of gluonic subsystem in Minkowski space is:
  \be
 \hat
   H_{gl}=\frac{p^2_r}{2\mu_g}+\frac{\hat{\vec{L}}^2_{gl}}
   {2(2\Pi+\mu_g)r^2}+ \int^1_0d\beta\nu(\beta)+\frac{\mu_g}{2}+
   \sigma^2r^2\int^1_0\frac{d\beta}{\nu(\beta)}
   \ee
    Here $\hat{\vec L}_{gl}$
   is the angular momentum of the gluon ; $p_r$
    is the radial momentum.  We believe that physical
   states of gluon are electric and magnetic ( see [1]). For lowest
   states $<\hat{\vec{L}}_{gl}^2>=2$.

     Here we
   are to use some variational procedure. We choose an anzats for the
   effective string energy $\nu(\beta)$
    and
   define free parameters from the minimization of gluon subsystem
   energy .  After that quark correction  with this ansatz is
   calculated from (11),(12) and (14) .  The simplest ansatzes for
   $\nu(\beta)$ are :
 $$
 \begin{array}{ll}
 \nu(\beta)= Const&(Ansatz I)\\
 \nu(\beta)= C_1+C_2(\beta-1)^2&(Ansatz II)\\
 \nu(\beta)= \sigma r &(Ansatz III)
 \end{array}
 $$

   Without Coulomb interaction it's
 worthwhile  to write out values of interest in  the dimensionless
 form :
 \be
 E_{gl}/\sqrt{\sigma}=4.30\pm 0.05;~~
 E_{q\bar q}/(\sqrt{\sigma}\sqrt{\sqrt{\sigma}/m})=1.4\pm 0.2
 \ee

  These values were
 obtained by means of trial radial wave function of the Gaussian type.
  Numerically gluon energy is about 1.7 GeV , quark correction
 value equals hundreds  of MeV.

 Three Coulomb terms should be added
 to the full Hamiltonian:
 \be
 V_{coul}=
 -\frac{3\alpha_s}{2}(\frac{1}{r_1}+\frac{1}{r_2})+\frac{1}{6}
 \frac{\alpha_s}{\rho}
 \ee
  Since $\alpha_s/6\ll 1$
 quark-antiquark repulsion gives only small contribution into the
 resulting  energy .

 But  the main problem as it  is usual for most
 of  the (potential-like) models  is to define absolute scale of mass.
   Some additional constants are needed to fit hadronic spestrum and
 these constants vary for different sectors  of the theory.
 Following the recipe described in [ 1 ] we use the constant for
 heavy-light sector multiplied by two ( cause we have  two strings
 vs.  one string for heavy-light meson , and one can expect that
 perimeter-type terms for Wilson loop or hadronic shifts  could be
 twice larger than for the only string).  The procedure used is :
1) to define constant  for heavy-light meson for given heavy quark
mass (light quark is massless) .  By means of variational
                calculations with  known masses of the heavy-light
                S-states  (1975 Mev  for D-meson and 5300 MeV  for
                B-meson  - accuracy more than tens of MeV is not
                needed) we define additive negative constant
                $C_0$.
2)The resulting mass is:
                \be
                M_{q\bar{q}g}=2m_q+E_{gl}+E_{q\bar q}+2 C_0
                \ee

              The main results of this paper are
            summarized in Table I .  There hybrid masses for various
            heavy quark masses are listed.  The parameters  are
           $\alpha_s=0.3;~~\sigma=0.18 GeV^2$
           .  In
            the last column results obtained  by method  of [ 1 ] are
            shown.

\newpage

\begin{center}
{\bf Table I.}

{\it   Calculated  hybrid  masses, GeV}

\end{center}
\begin{tabular}{llllll}
&&   Ansatz I&   Ansatz  II&   Ansatz  III&   Potential\\
$m_c=$&1.2  GeV&        4.008&        3.951&        3.873&
4.089\\
   &    1.3&        4.004&        3.95&        3.875&         4.082\\
    &   1.5&        3.997&        3.946&        3.878&         4.069\\
     &  1.7&        3.991&        3.943&        3.88&         4.058\\
$m_b$= &4.2&       10.828&       10.810&       10.647&
10.647\\
& 4.5&       10.551&       10.521&       10.491&
       10.585\\
       &4.9&       10.55&       10.521&       10.492&
       10.583\\
       &5.2&       10.548&       10.519&       10.491&
       10.58 \end{tabular}

             From  Table I  one  can  conclude  that  our  model
             anticipate  for  c.o.m. of  hybrid  multiplet
\be
M=\left \{
\begin{array}{lll}
4.0\pm 0.1 GeV& for& c\bar c g\\
10.5\pm 0.1 GeV& for& b\bar b g
\end{array}
\right.
\ee
          Quantum  numbers  of  hybrids  are  given  in  [ 1 ] .

              Spin-dependent  forces   are  not  considered  in  this
          paper ( the  most important  are $(\vec L\vec S)$.  They  will  be
discussed  in  the
          forthcoming studies.

            The  result  (19)  is  in a  good
          agreement with  that  ones  of  [3]  and  [4].
          Unfortunately, this is  a  rather  accident  fact  cause
          in  [3], [4]  and in  this  paper  various  prescriptions
          for  additive constants  in  the  interquark  potential
          were   used.  Thus  we  believe   that  there  is  no
          reasons  for exceed  optimism  and  additional  attempts
          should  be made  to  understand  the  origin  of  the
          constant  term.

 This research was made possible by Grant N J77100 from the International
 Science Foundation and Russian Government.

 The support from Russian Fundamental Research Foundation,Grant N
 93-02-14937 is also acknoledged.


\newpage

\begin{thebibliography}{99}
\bibitem {1}
 Yu.S.Kalashnikova, Yu.B.Yufryakov, Preprint
ITEP-35-95 , hep-ph/9506269 ( to  be  published  in  Physics  Letters
B ).
\bibitem {2}
   A.Yu.Dubin, A.B.Kaidalov  and Yu.A.Simonov,Phys. Lett.
  B323,41(1994); E.L.Gubankova and A.Yu.Dubin, Phys.Lett.
  B334,180(1994)
\bibitem {3}
        J.Merlin and  J.Paton , J. Phys. G11 , 439
  (1985) N. Isgur  and  J.Paton, Phys. Rev. D31, 2910 (1985)
       J.Merlin  and  J.Paton, Phys. Rev. D35, 1668 (1987)
\bibitem {4}
   S.Perantonis  and C.Michael, Nucl. Phys. B347, 854 (1990)
\bibitem {5}
   T.Barnes , F.E.Close  and  E.S. Swanson , RAL-94-106, hep-ph/9501405
   \end{thebibliography}
   \end{document}